# Illusory versus Genuine Control in Agent-Based Games

J.B. Satinover[1,a] and D. Sornette[2,b]

[1]Laboratoire de Physique de la Matière Condensée, CNRS UMR6622 and Université des Sciences, Parc Valrose, 06108 Nice Cedex 2, France

[2]Department of Management, Technology and Economics, ETH Zurich, CH-8032 Zurich, Switzerland

**Abstract:** In the Minority, Majority and Dollar Games (MG, MAJG, $G) synthetic agents compete for rewards, at each time-step acting in accord with the previously best-performing of their limited sets of strategies. Different components and/or aspects of real-world financial markets are modelled by these games. In the MG, agents compete for scarce resources; in the MAJG agents imitate the group in the hope of exploiting a trend; in the $G agents attempt to successfully predict and benefit from trends as well as changes in the direction of a market. It has been previously shown that in the MG for a reasonable number of preliminary time steps preceding equilibrium (Time Horizon MG, THMG), agents' attempt to optimize their gains by active strategy selection is "illusory": The calculated hypothetical gains of their individual strategies is greater on average than agents' actual average gains. Furthermore, if a small proportion of agents deliberately choose and act in accord with their seemingly worst performing strategy, these outperform all other agents on average, and even attain mean positive gain, otherwise rare for agents in the MG. This latter phenomenon raises the question as to how well the optimization procedure works in the MAJG and $G. We demonstrate that the illusion of control is absent in MAJG and $G. In other words, low-entropy (more informative) strategies under-perform high-entropy (or random) strategies in the MG but outperform them in the MAJG and $G. This provides further clarification of the kinds of situations subject to genuine control, and those not, in set-ups a priori defined to emphasize the importance of optimization.

PACS. 89.75.-k Complex systems – 89.65Gh Economics; econophysics, financial markets, business and management – 02.50.Le Decision theory and game theory

## 1 Introduction

Langer's phrase, "illusion of control" [1] describes the fact that individuals appear hard-wired to over-attribute success to skill, and to underestimate the role of chance, when both are in fact present. We have previously shown [2] that in one of the most extensively-studied agent-based game-model of markets, the minority game (MG), agents' control is illusory in the following sense: The mean actual performance of all agents averaged over many different initial conditions is poorer than the mean hypothetical performance of all their given strategies. The finding is striking because at each time-step agents deploy that strategy with the best such hypothetical performance. This finding is most generally true under the following conditions: The initial state is iterated for a "reasonable" number of time-steps ($\leq 2000$ say) short of equilibrium ($\geq 5000$ say) at which point a rolling window of cumulative strategy scores is maintained, which defines the specificity of the Time Horizon MG (THMG) as compared with the MG. In this case, the illusion is observed for all m (m is the number of bits of binary history in agents' memory and an "m-bit history" consists of a string of 1's and 0's of length m. 1 indicates that a minority of agents have adopted action "-1", 0 indicates that a minority of agents have adopted action "+1"). The finding is less generally true in the MG strictly speaking, in which the system is always allowed to run to equilibrium. In that case, it is true only for almost all m, and not true (but only, and just barely) at the so-called critical point $\alpha_c = \frac{2^{m_c}}{N} = 0.34$, where $\alpha \equiv \frac{2^m}{N}$ in general and N is the number of agents. For example with N=31, $3 < m_c = 3.4 < 4$. Effectively, this means that $m_c \equiv 4$ as memory is in discrete units. However, equilibrium in the MG proper is only reached after $\sim 100 \times 2^{m+1}$ iterations away from $\alpha_c$ and orders or magnitude more near $\alpha_c$, therefore for N=31 at $m_c$=4, $t_{eq} \gg 3200$ iterations. Arguably, no real-world market is sufficiently stationary to attain such an equilibrium state.

---
[a] jsatinov@princeton.edu
[b] dsornette@ethz.ch

Another important finding presented in [2] is the surprising fact that, if a small proportion of agents (≤ 0.15, e.g., 3 of 31) deploy at each time-step their previously *worst* performing strategy, these agents on average outperform all others. In fact they relatively consistently attain on average net positive gain, which is otherwise rare in the MG as the mean gain for both strategies and agents is in general always negative due to the minority rule. Note that such an inversion of the selection rule is a symmetric alteration in agent behavior and involves no privileging or increase in computational capacity.

The success of such "counteradaptive" agents is a further marker of the "illusory" nature of the standard optimization rule ("choose best"). But it also raises the following question: May one correctly think of such an inverted rule as equivalent to these agents' playing a Majority Game (MAJG) instead? It would seem on the face of it that as they are not optimizing to be in the minority, they must be optimizing to be in the majority but failing, and rather inadvertently succeed in finding the minority remarkably often. By this reasoning it seems to follow that in a game where all agents are striving to be in the majority (MAJG), select agents that optimize instead to be in the minority will likewise succeed disproportionately implying that the MAJG should also demonstrate an "illusion of control". A similar argument could be made, perhaps, with respect to the $G since players of this game are rewarded according to a time-lagged majority rule. (The formal distinctions among the three types of games are specified in the following section.)

The goal of the present paper is to clarify these questions and demonstrate that agents who invert their optimization rule in the MG are not actually playing a MAJG and that no illusion of control is found in either the MAJG or the $G. We discuss our results in terms of persistent versus anti-persistent characteristics of the time series generated by the various models. In a follow-up to this paper [3], we relate these comparative results to different characteristics of markets—or different phases they may enter—as real-world agents alter their game-playing behavior.

We first briefly review the formal structure of each of the MG, MAJG and $G. We then present the results of extensive numerical simulations. Finally we discuss the differences that emerge in terms of persistent versus anti-persistent time-series.

## 2 Minority, Majority and $ Games

### 2.1 Definition and overview of Minority Games (MG) and Time-Horizon MG (THMG)

MGs exemplify situations in which the "rational expectations" mechanism of standard economic theory fails. This mechanism in effect asks, "what expectation model would lead to collective actions that would on average validate the model, assuming everyone adopted it?"[4]. In minority games, a large number of interacting decision-making agents, each aiming for personal gain in an artificial universe with scarce resources, try to anticipate the actions of others on the basis of incomplete information. Those who subsequently find themselves in the minority group gain. Therefore, expectations that are held in common negate themselves, leading to anti-persistent behavior both for the aggregate behavior and for individuals. Minority games have been much studied as repeated games with expectation indeterminacy, multiple equilibria and inductive optimization behavior [5,6].

Each of the N players have to choose one out of two alternatives at each time step based on information represented as a binary time series A(t). Those who happen to be in the minority win. Each element of the time series encodes which alternative is the winning (minority) one. Each agent is endowed with S strategies. Each strategy gives a prediction for the next outcome A(t) based on the history of the last m realizations A(t-1), …, A(t-m) (m is called the memory size of the agents). Each agent holds the same number S of (in general different) strategies among the $2^{2^m}$ total number of strategies. The S strategies of each agent are chosen at random but once and for all at the beginning of the game—thus the system is quenched at its initial disorder. At each time t, in the absence of better information, in order to decide between the two alternatives for A(t), each agent uses her most successful strategy in terms of payoff accumulated since t=1. If instead hypothetical points are summed over a rolling window of finite length $\tau$ up to the last information available at the present time t, the game is now the "Time Horizon MG" (THMG; the case of a limitlessly growing $\tau$ corresponds to the standard MG; the THMG refers to the case of a fixed and finite $\tau$). This is the key optimization step. If her best strategy predicts A(t)=+1 (resp. -1), she will take the action $a_i(t)$ = - 1 (resp. +1) in order to hope to be in the minority. The aggregate behavior A(t) = $\sum_{i=1}^{N} a_i(t)$ is then added to the information set available for the next iteration at time t+1.

The characteristic of the MG that distinguishes it from the MAJG and the \$G is the corresponding instantaneous payoff of agent i which is given by – $a_i(t)$ A(t)—the minus sign encodes the minority rule (and similarly for each strategy for which it is added to the $\tau$-1 previous payoffs. More simply, the payoff may also be – Sign[$a_i(t)$ A(t)]. This does not change the fundamental dynamics of the game nor has it any effect on the question we are here studying). As the name of the game indicates, if a strategy is in the minority ($a_i(t)$A(t) < 0), it is rewarded. In other words, agents in the MG and THMG try to be anti-imitative.

The richness and complexity of minority games stem from the fact that agents strive to be different. Previous investigations have shown the existence of a phase transition marked by agent cooperation and efficiency between an inefficient regime (worse than random) and a random-like regime as the control parameter $\alpha \equiv 2^m/N$ is increased: In the vicinity of the phase transition at $\alpha_c = 2^{m_c}/N \approx 0.34$ (for both the THMG and MG proper), the size of the fluctuations of A(t) (as measured by its normalized variance $\sigma^2/N$) falls below the random coin-toss limit for large *m*'s (assuming fixed N) when agents always use their highest scoring strategy [5]. In other words, for a range of m (given N, S), agent performance is better than what strategy performance would be in a game with no agents optimizing. The phenomenon discussed in [2] is that when optimizing, and averaged over all actual agents and their component strategies in a given realization, and then averaged over many such initial quenched disorder states, agents in the TH variant of the MG nonetheless underperform the mean of their own measured strategy performance and do so in all phases for reasonable lengths of $\tau$ at all m. (In the MG, the same statement holds true for "reasonable" run lengths post initialization but pre-equilibrium. It holds true post-equilibrium as well except for m at or very near $\alpha_c$, but in this region the number of steps to equilibrium is extremely large).

More specifically, for any given realization, a minority of agents outperform their strategies and the majority of other agents. (Some may also achieve net positive gain, if rarely.) In the MG proper, however, $\tau$ is unbounded and a stationary state is reached at some very large $\tau_{eq} \geq 2^m \times 200$ where a subset of agents "freeze" their choice of strategy: One virtual strategy score attains a permanently higher value than any other. These frozen agents in general do outperform the mean of all strategies in a given realization as well as the mean of their own S original strategies: They perform precisely as well as their best. We focus primarily on results in the THMG with an eye towards real-world markets in which because the time series being predicted are non-stationary, trading strategies are weakened if they incorporate an unbounded (and uniformly-weighted) history of prior strategic success or failure: Remote history is less important than recent history and beyond a certain point is meaningless. Unless specifically stated otherwise, throughout this paper, whenever we compare agent to strategy performance, we always mean the performance of agents' strategies as measured by the accumulation of hypothetical points averaged over *all* agents in the system and the set of *all* of their strategies. Furthermore, in selecting a strategy, the agents do not take account of the impact of their choice on the probable minority state—that is, they do not consider that their own selection of action reduces the probability that this action will be the minority one. (We refer to such agents as "standard".)

## 2.2 Definition and overview of the Majority Game (MAJG)

Mathematically, the MAJG game differs from the MG (and a THMAJG from the THMG) only by a change in sign for the individual agent payoff function: i.e., $g_i^{min}(t) = -a_i(t)A(t)$ or $g_i^{min}(t) = -Sgn[a_i(t)A(t)]$ whereas $g_i^{maj}(t) = +a_i(t)A(t)$ or $g_i^{maj}(t) = +Sgn[a_i(t)A(t)]$. In consequence of the plus sign, agents are rewarded when they select the alternative selected by the majority of agents at a time step. Thus, agents strive to be imitative rather than anti-imitative. From the perspective of markets, agents in the MG are "pessimistic" in assuming that resources are limited so that there can be only a minority of winners; they are "contrarian" in attempting to do what they believe most others are not doing. Agents in the MAJG are "optimistic" in assuming that resources are boundless, price (and value) potentially rising simply by virtue of collective agreement, so that the majority wins; they are "conformist" in attempting to do what they believe most others are also doing. Agents in both types of games "believe" that their actions may be optimized by examining the past paper-performance of their strategies.

As only a minority of agents win in the MG, mean agent gain $\bar{G}^{min}(t) = \frac{1}{N}\sum_{i=1}^{N} g_i^{min}(t) < 0$. Cumulative wealth tends to decrease over time. In the MAJG, a majority of agents win so that mean agent gain $\bar{G}^{maj}(t) = \frac{1}{N}\sum_{i=1}^{N} g_i^{maj}(t) > 0$. Cumulative wealth tends to increase over time.

In the MG, the time series A(t) (the sum of all agents' actions) is typically anti-persistent, paralleling the anti-imitative behavior of individual agents. In the MAJG the time series A(t) is typically persistent, paralleling the imitative behavior of individual agents.

## 2.3 Definition and overview of the Dollar Game ($G)

The dollar game was introduced in [8] in order to capture more accurately the behavior of traders in markets, while keeping a framework as close as possible to the initial MG set-up. Above we commented that agents in the MG are "contrarian" in attempting to do what most others are not. But this is not what true contrarian traders attempt: First, they attempt to be in the majority when the market is rising. They likewise attempt to be in the minority when it is falling or when there is a turning point. And this is exactly what non-contrarian traders are also attempting. Indeed every trader attempts to do this. Contrarians differ from conformists in their *reasoning* as to what the market trend will be in the immediate future. They make predictions that typically differ from the majorities' prediction—but they may or may not be correct. Like all others, they will still hope that, if correct, it will lead them to be in the majority in one instance and the minority in the other, as is appropriate according to the corresponding market phase. A similar correction to the description of "conformist" traders can be made.

Thus, an agent with greater "real world" behaviors is precisely one that rationally alternates between choosing what he believes will be the minority state and choosing what he believes will be the majority state. Ideally, he wants to start choosing to try to be in the majority state at the first moment the market begins a rise following a decline—i.e., at a convex inflection point. Likewise, he ideally wants to start choosing to try to be in the minority state at the first moment the market begins a decline following a rise—i.e., at a concave inflection point. This behavior may be most simply captured by the following alteration in the rule for individual agent gain: $g_i^\$(t) = +a_i(t-1)A(t)$ or $g_i^\$(t) = +Sgn[a_i(t-1)A(t)]$. That is, it is the action at the previous time step t-1, interpreted as a judgment about whether A(t) will be >0 or <0, that determines whether an agent gains or loses. The mean agent gain retains the same form: $\overline{G}^\$(t) = \frac{1}{N}\sum_{i=1}^{N} g_i^\$(t)$ and we anticipate that $\overline{G}^\$(t) > 0$ because in spite of the time-lagged $a_i(t-1)$, the payoff function is preceded by a + sign, so intuitively should generate largely imitative behavior. This intuition is confirmed by the numerical simulations presented in Ref.[8].

## 3. Main results on the "illusion of control" in the THMG v. MAJG and $G

Our main result with respect to the THMG may be stated concisely from the perspective of utility theory and has been detailed extensively in [2]: Throughout the space of parameters (N, m, S, $\tau \ll \tau_{eq}$), the mean payoff of agents' strategies (as calculated by each agent averaged over all strategies and agents in a realization) not only surpasses the mean payoff of supposedly-optimizing agents (averaged over all given agents), but the respective cumulative distribution functions (CDF) of payoffs show a first-order stochastic dominance of strategies over agents. Thus, were the option available to them, agents would behave in a risk-averse fashion (concave utility function) by switching randomly between strategies rather than optimizing. Agents are supposed to enhance their performance by choosing adaptively between their available strategies. In fact, the opposite is true. (In the MG proper the situation is more complex and is discussed at length in [2]. Here we note only that two conditions must hold for the statement to be false: (1) m ≥ $m_c$; (2) the system must be allowed to reach equilibrium. Condition (2) requires an exceedingly large number of preliminary steps before agent selection begins, and orders of magnitude more steps if m ≈ $m_c$ ).

Let us restate our result for the THMG in the language of a financial market with traders trying to outperform the overall market. We argue that, in using the THMG as a model for traders' actions, the following is the case: Every trader attempting to optimize by selecting his "best performing strategy" measures that performance virtually, not by contrast to an imagined setting where all traders select fixed strategies at random (to whose results he would have no access anyway). Even though the virtual performances of each of his basket of strategies might never have been implemented in reality, if he found that his real performance under a selection process was worse than the virtual performance of the strategies he had been selecting among, he would abandon the selection process. This would be true for most agents and not true only for a small minority. (If *every* trader were to do the same, of course, then one would end up with a random or fixed choice game in which agents' mean performance approaches $\sigma^2/N = 0.5$, where $\sigma$ is the standard deviation of the variation in A(t). (This forms the usual standard of comparison for strategy performance in the MG literature.) This resonates with the finding of Doran and Wright [9], who report that two-thirds of all finance professors at accredited, four-year universities and colleges in the U.S. (arguably among the most sophisticated and informed financial investors) are passive investors who think that the traditional valuation techniques are all *un*important in the decision of whether to buy or sell a specific stock (in particular, the CAPM, APT and Fama and French and Carhart models).

However, in both the MAJG and the $G, we find that the reverse is true: The optimization method greatly enhances agent performance, with strategies' virtual mean performance consisting of relatively small gains and agents' mean performance consisting of significantly greater gains. In the language used above: "Throughout the space of parameters (N, m, S, $\tau \ll \tau_{eq}$), the mean payoff of agents' strategies (as calculated by each agent averaged over all strategies and agents in a realization) not only underperforms the mean payoff of optimizing agents (averaged over all given agents), but the respective cumulative distribution functions (CDF) of payoffs show a first-order stochastic dominance of agents over strategies. Thus, were the option available to agents to behave in a risk-averse fashion (concave utility function) by switching randomly between strategies rather than optimizing, they would rationally avoid such risk in favor of the optimization procedure. Agents are supposed to enhance their performance by choosing adaptively between their available strategies and they in fact do so."

## 4. Quantitative statement and tests

### 4.1 Analytic Calculation versus Numeric Simulation

In the THMG, the "illusion of control" effect is observed for all N, m, S and $\tau \ll \tau_{eq}$. We use the Markov chain formalism for the THMG [7] and extend it to both a THMAJG and a TH$G to obtain theoretical prediction for the gains, $\Delta W_{Agent}$ averaged over all agents and $\Delta W_{Strategy}$ averaged over all strategies respectively, of agents and of all strategies in a given realization [10] for each of the MG, MAJG and $G:

$$\left\langle \Delta W_{agent}^{game} \right\rangle = \pm \frac{1}{N} \left| \vec{A}_D \right| \cdot \vec{\mu} \tag{1}$$

$$\left\langle \Delta W_{strategy}^{game} \right\rangle = \frac{1}{2N} \left( \hat{\mathbf{s}}_\mu \cdot \vec{\kappa} \right) \cdot \vec{\mu} \tag{2}$$

In Eq.s (1) and (2), brackets denote a time average. The superscript "*game*" identifies the game type with $game \in \{M, MAJ, \$\}$. In Eq. (1), the minus sign is needed for the MG; otherwise not. $\mu$ is a $(m+\tau)$-bit "path history" [10] (sequence of 1-bit states); $\bar{\mu}$ is the normalized steady-state probability vector for the history-dependent $(m+\tau) \times (m+\tau)$ transition matrix $\hat{\mathbf{T}}$, where a given element $T_{\mu_t, \mu_{t-1}}$ represents the transition probability that $\mu_{t-1}$ will be followed by $\mu_t$; $\bar{A}_D$ is a $2^{(m+\tau)}$-element vector listing the particular sum of decided values of A(t) associated with each path-history; $\hat{\mathbf{s}}_\mu$ is the table of points accumulated by each strategy for each path-history; $\bar{\kappa}$ is a $2^{(m+\tau)}$-element vector listing the total number of times each strategy is represented in the collection of N agents. As shown Ref. [2], $\hat{\mathbf{T}}$ may be derived from $\bar{A}_D$, $\hat{\mathbf{s}}_\mu$ and $\bar{N}_U$, the number of undecided agents associated with each path history. Thus agents' mean gain is determined by the non-stochastic contribution to A(t) weighted by the probability of the possible path histories. This is because the stochastic contribution for each path history is binomially distributed about the determined contribution. Strategies' mean gain is determined by the change in points associated with each strategy over each path-history weighted by the probability of that path.

Agreement is excellent between numerical simulations and the analytical predictions (2.1) and (2.2) for the THMG, THMAJG and TH$G. For instance, for m=2, S=2, τ=1 and N=31, for one identical quenched disorder state, **Table 1** shows the payoff per time step averaged over time and over all agents and all strategies for both analytic and numerical methods. In this numerical example, the average payoff of individual agents is smaller than for strategies by −0.15 units per time step in the THMG, but larger by +0.35 units in the THMAJG and by +0.33 units in the TH$G. Thus, in this example, optimization appears to agents as genuine in the THMAJG and TH$G but would seem illusory in the THMG.

**Table 1**: Numeric and Analytic Results for a single typical quenched initial disorder state in the THMG, THMAJG and TH$G

| Numeric | $\langle \Delta W_{agent}^{game} \rangle$ | $\langle \Delta W_{strategy}^{game} \rangle$ | Analytic | $\langle \Delta W_{agent}^{game} \rangle$ | $\langle \Delta W_{strategy}^{game} \rangle$ |
|---|---|---|---|---|---|
| MG | −0.21 | −0.06 | MG | −0.21 | −0.06 |
| MAJG | +0.43 | +0.08 | MAJG | +0.43 | +0.08 |
| $G | +0.39 | +0.06 | $G | +0.40 | +0.06 |

The above results illustrate primarily the close alignment of analytic and numerical methods in generating results. Of greater interest is the comparison of agent versus strategy gains among the MG, MAJG and $G at various values of m below, at and above $m_c$, and at various values of τ both for $\tau \ll \tau_{eq}$, for $\tau < \tau_{eq}$ and for $\tau \geq \tau_{eq}$—all averaged over a large ensemble of randomly selected quenched disorder states. The computational resources required to evaluate the analytic expressions grows for $\langle \Delta W_{agent}^{game} \rangle$ as $\propto 2^{m+\tau}$. We therefore report only the numerical results.

### 4.2 Illusory versus Genuine Control for $\tau \ll \tau_{eq}$

Almost all results that hold for multiple values of $\tau \ll \tau_{eq}$ are illustrated for $\tau = 1$. In **Figure 1**, **Figure 2** and **Figure 3** we therefore first present graphic representations of the ensemble average of 50 runs comparable to **Table 1** but over many values of m.

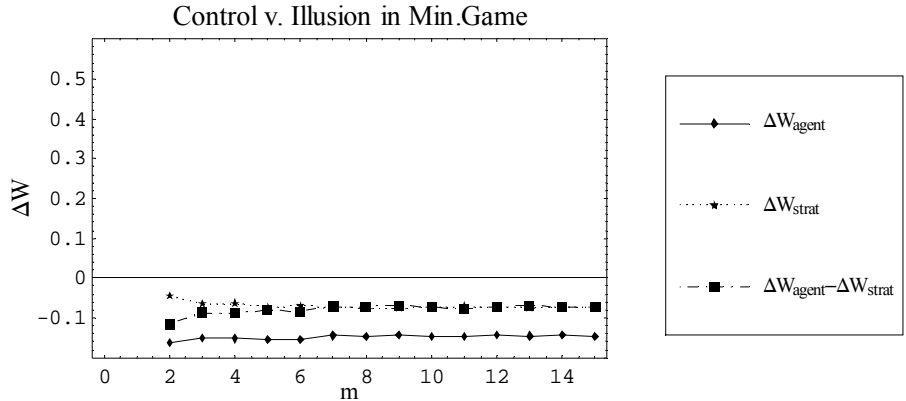

**Figure 1**: Agent versus Strategy mean per-step gain in the THMG at various m with τ=1. The phase transition expected at m=4 is absent; strategies outperform agents at all m as indicated by the black square: Agent performance is always negative relative to strategy performance. The optimization procedure employed by agents yields worse performance than their component strategies on the basis of which agents select which strategies to deploy at each time-step.

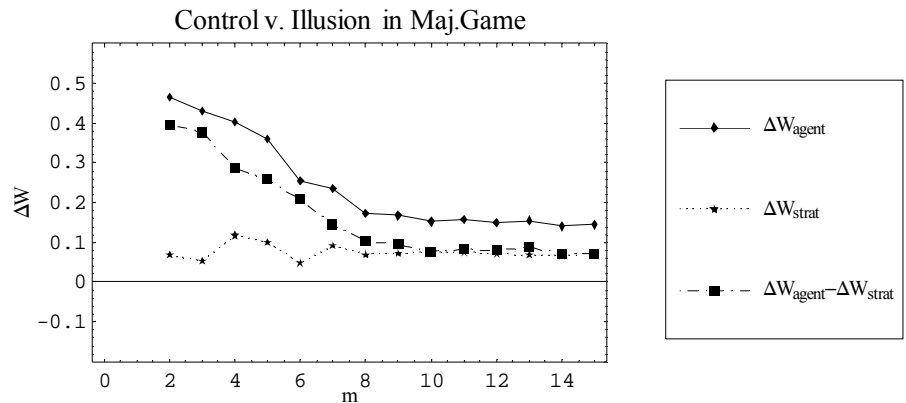

**Figure 2**: Agent versus Strategy mean per-step gain in the THMAJG at various m with τ=1. Agent performance is always positive and greater than strategy performance. The optimization procedure employed by agents yields better performance than their component strategies.

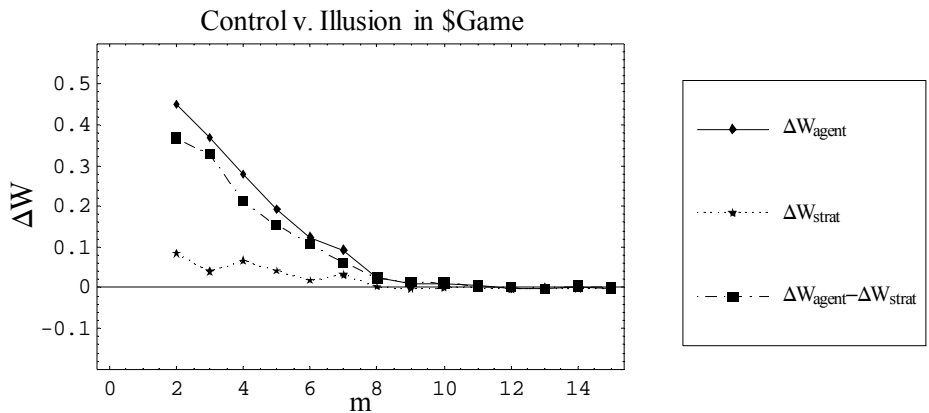

**Figure 3**: Agent versus Strategy mean per-step gain in the TH$G at various m with $\tau=1$. Agent performance is always greater than strategy performance. The optimization procedure employed by agents yields better performance than their component strategies, but the gain becomes asymptotically small for large m's.

We see that the illusion of control in the THMG persists at all values of m. Incidentally we note that the phase transition to be expected at m=4 is strongly surpressed in the sense that the present metric is not sensitive to it. For both the THMAJG and the TH$G, the control exerted by agents is non-illusory: Agents outperform their constituent strategies at all m. Because of the non time-lagged implementation of a majority rule in the THMAJG, strategies show consistent positive gain, even if less than agents. Strategies' gain tends toward a positive limit with agents' gain tending toward a greater value at all m. However, in the TH$G, strategies on their own, in the aggregate, tend toward zero gain with increasing m, as would be expected from a realistic model of a market. Agents are superior to strategies at all m, but converge to the zero limit of strategy gain with increasing m. In other words, of the three variations, the TH$G with very short $\tau$ shows the most satisfying convergence toward neither net positive nor net negative gain for both strategies and agents as strategy complexity increases and begins to approximate random selection. It is especially interesting that this is so, given that the $G rule remains a majority one, albeit time-lagged by one step to take into account the time lag between decision and return realization [8].

### 4.3 Illusory versus Genuine Control for $\tau < \tau_{eq}$

In **Figure 4**, **Figure 5** and **Figure 6** we present graphic representations of the ensemble average of 50 runs of the MG, MAJG and $G comparable to **Table 1** but over many values of m and with $\tau < \tau_{eq}$, i.e. with a time window of "reasonable" size, but smaller than the equilibrium value (except for m=2, where $\tau_{eq} = 800$).

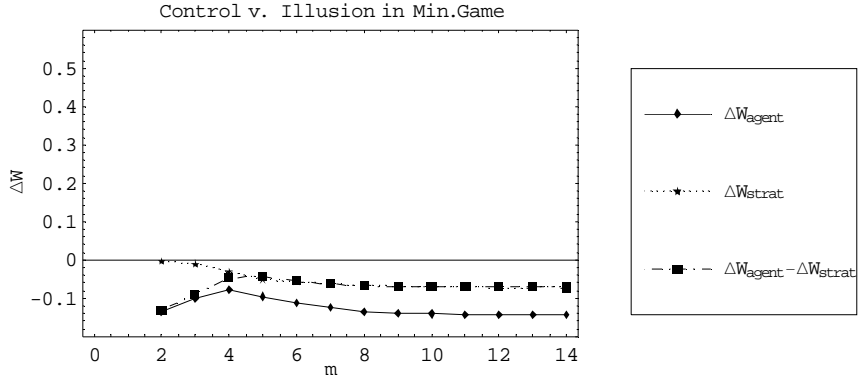

**Figure 4**: Agent versus Strategy mean gain per-step in the THMG at various m with τ=1000. The phase transition expected at m=4 is clearly visible; strategies outperform agents at all m as indicated by the black square: Agent performance is always negative relative to strategy performance. Even with a very long lookback of historical data, the optimization procedure employed by agents yields worse performance than their component strategies on the basis of which agents select which strategies to deploy at each time-step.

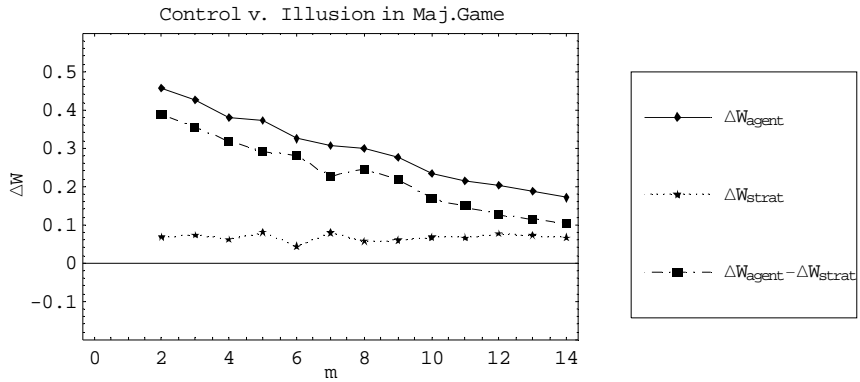

**Figure 5**: Agent versus Strategy mean gain per-step in the THMAJG at various m with τ=1000. Agent performance is always positive and greater than strategy performance. The optimization procedure employed by agents yields better performance than their component strategies.

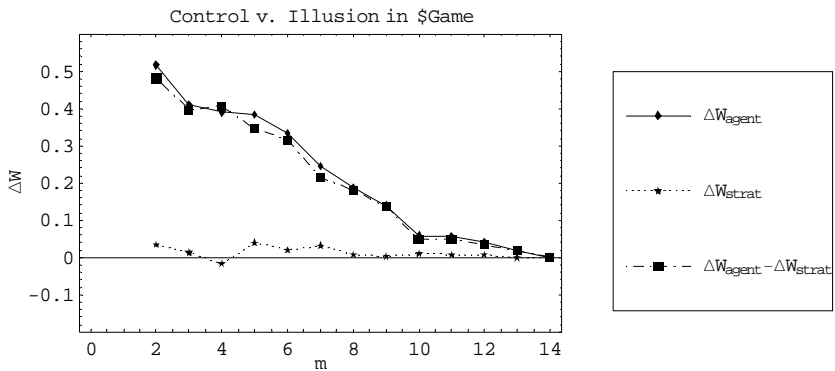

**Figure 6**: Agent versus Strategy mean gain per-step in the TH$G at various m with τ=1000. Agent performance is always greater than strategy performance. The optimization procedure employed by agents yields better performance than their component strategies.

We see that once again, the illusion of control in the THMG persists at all values of m in spite of the "reasonable" length (1000 time steps) of τ. Note, however, that the phase transition at m=4 is now visible—even though at m=4 the system is still far from equilibrium. (Recall that away from $m_c$, we have $\tau_{eq} \approx 100 \times 2^m$; while for m near $m_c$, we have $\tau_{eq} \gg 100 \times 2^m$.) In the MG proper, where τ grows without bound and agent and strategy performance begins to be measured only after $\tau_{eq}$ steps, agent performance will exceed strategy performance—optimization succeeds—but only for $m \geq m_c$. But even for a relatively small number of agents (e.g., 31, as here), at m=10 say, $\tau_{eq} \approx 100 \times 2^{11} > 200,000$ steps is unrealistically large (for a comparison with standard technical investment strategies used for financial investments). For both the THMAJG and the TH$G, the control exerted by agents is again non-illusory: Agents outperform their constituent strategies at all m. Strategies in the THMAJG again show consistent positive gain, if less than agents. Strategies' gain likewise tends toward a positive limit with agents' gain tending toward a greater value at all m, just as for $\tau = 1$. Likewise in the TH$G once more: Strategies on their own, in the aggregate, tend toward zero gain with increasing m, as would be expected from a realistic model of a market. Agents are superior to strategies at all m, but converge to the zero limit of strategy gain with increasing m. We may draw a similar conclusion for $\tau = 1000$ as for $\tau = 1$: The TH$G with reasonable τ shows the most satisfying convergence toward neither net positive nor net negative gain for both strategies and agents as strategy complexity increases and begins to approximate random selection, in line with what would be expected from the efficient market hypothesis [12,13].

## 5. Interpretations: crowding-out, anti-optimizing agents and persistence

### 5.1 Illusion of control and the crowding-out mechanism

Illusion-of-control effects in the THMG result from the fact that a strategy that has performed well in the past becomes crowded out in the future due to the minority mechanism [2]: Performing well in the recent past, there is a larger probability for a strategy to be chosen by an increasing number of agents, which inevitably leads to its failing. Optimizing agents tend on average to adapt to the past but not the present. They choose an action a(t) which is on average out-of-phase with the collective action A(t). In contrast, non-optimizing (random- or fixed-choice) agents average over all the regimes for which their strategy may be good and bad, and do not face the crowding-out effect. The crowding-out effect also explains simply why anti-optimizing agents over-perform [2]: Choosing their worst strategy ensures that it will be the least used by other agents in the next time step, which makes it more probable that they will be in the minority. In Ref.[2], we consider generalizations of the MG in which more clever agents use this insight to develop strategies based on their awareness of prior-level effects: when agents use a boundless recursion scheme to learn and optimize their strategy, the game converges to an equilibrium with fully symmetric mixed strategies where agents randomize their choice at each time step with unbiased coin tosses! The crowding mechanism also predicts that the smaller the parameter $2^m/N$, the larger the illusion-of-control effect. Indeed, as one considers larger and larger values of $2^m/N$, it becomes more and more probable that agents have their strategies in different reduced strategy classes, so that a strategy which is best for an agent tells nothing about the strategies used by the other agents, and the crowding out mechanism does not operate. Thus, regions of successful optimization, if they occur at all, are more likely at higher values of $2^m/N$ (See Appendix A for further details.)

By contrast, in all of the MAJG, $G, THMAJG and TH$G, with their variants of a majority mechanism for agent gain, a strategy that has performed well in the past is likely to do so again in the future. The domain of successful optimization encompasses all m, but diminishing as m increases and strategies become widely dispersed in strategy space, approximating ever more closely a collection of random decision makers. The optimization procedure is most effective in the $G where the positive bias present even for strategies alone in the MAJG appears neutralized by the time-delayed factor: On their own, strategies show effectively neither gain nor loss. Gains are therefore due solely to the optimization procedure. Given this, we predict that anti-optimizing agents should show no advantage over their optimizing counterparts in the MAJG and $G and will rather underperform. The next section presents results of simulations testing this prediction.

**5.2. Illusion of control and "anti-optimizing" agents**

We select 3 of 31 agents to function "counteradaptively" ("c agents") and the remaining to function in the standard fashion ("s agents"). c-agents "anti-optimize"—at each time-step they deploy that strategy with the *fewest* virtual points accumulated over τ, rather than the strategy with the most points as do s-agents. In [2], we studied this phenomenon in the "crowded regime" ($m < m_c, \alpha < \alpha_c$) where crowd and anti-crowd formation is likely and the minority mechanism causes the larger-sized crowds to lose. In this regime, c-agents that choose their worst performing strategy consistently outperform s-agents that choose their best. Here we display results obtained for a wide range of m both less than, and greater than $m_c$, for the THMG, THMAJG and TH$G with $\tau = 100$. τ is long enough so that the phase transition in the MG is not suppressed at $m_c$ (=4 for n=31).

**Figure 7** shows c-agent minus s-agent mean per-step change in wealth for 2<m<14, each averaged over 100 runs of 100 days post-τ=400. In the THMG, in the crowded regime, the illusion of control effect is so strong that c-agents significantly outperform s-agents. Because we know that, for all m at this τ, agents underperform strategies, we see that the opposite is true for c-agents: In the act of "anti-optimizing", they actually optimize. However, as the phase transition approaches, this becomes less true. Indeed at $m_c$ and after $m_c$—that is, in the non-crowded regime—s-agents outperform c-agents, converging to zero difference with increasing m. We know however that these s-agents for large m are nonetheless underperforming their strategies. Thus, while the illusion of control effect remains present, it is not strong enough for c-agents to outperform s-agents in this regime.

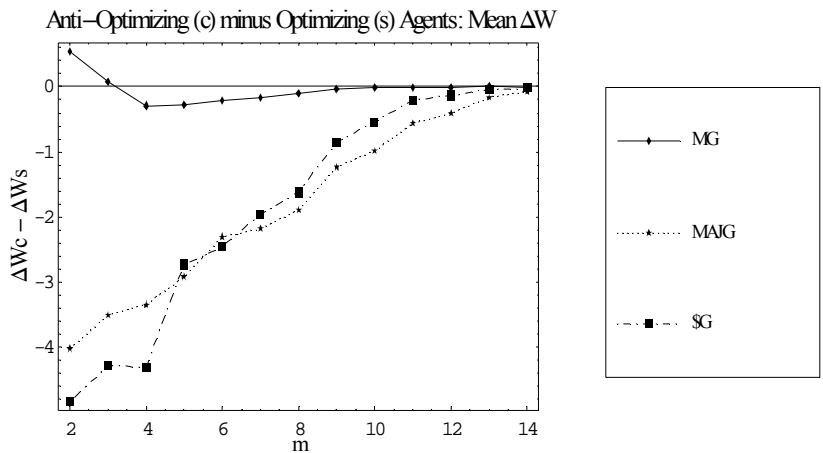

**Figure 7**: Difference between c-agent and s-agent mean per-step change in wealth for 3 of 31 c-agents, averaged over 100 days and 200 runs with τ=400 in the THMG, THMAJG and TH$G.

By contrast with the results for the THMG, c-agents in the THMAJG and TH$G consistently underperform s-agents as predicted from the success of the optimization scheme at all m (again converging to zero difference at large m). The size of this underperformance for anti-optimizing agents is consistent with the large degree of standard optimization success as shown in **Figure 5** and **Figure 6**.

### 5.3 Persistence versus Anti-persistence in the THMG, THMAJG and TH$G

As discussed in [2], in the MG and THMG, the degree to which agents underperform their own strategies varies with the phase as parameterized by $\alpha$. As noted in [7], in the crowded phase ($\alpha < \alpha_c, m < m_c$), the "crowd" of such agents choosing an action at any given time-step acts like a single "super-agent"; the remaining agents as a (non-synchronized) "anti-crowd" whose actions will conform to the minority choice. Thus, when a strategy is used, it is probably used by more than one agent, often by many agents. By enough, it becomes a losing strategy with large probability—precisely because so many agents "think" it's the best choice and use it. This implies that at the next time step, agents will not use it. The time-series of determined choices $\bar{A}_D$ therefore does not show trends (or persistence), but rather anti-persistence.

Anti-persistence is not equivalent to "random" and is scale-dependent. Consider a binary time-series with an *m*-bit $\mu(t)$ defined in the same way as we have in the MG or THMG: $\mu(t)$ is a sliding window of 1-bit states each of length *m*: $s(t-m), s(t-m+1), \ldots s(t)$. A *perfectly* anti-persistent binary series at scale *m*=2, for example, is characterized as follows: Select any one instance of the four possible $\mu(t) \in \{00, 01, 10, 11\}$. Identify the following bit $s(t+1) \in \{0,1\}$. Now identify the next instance of the selected $\mu(t)$. If the series is perfectly anti-persistent, the following bit will *always* be 1 if the previous following bit was 0, and 0 if the previous following bit was 1. A perfectly anti-persistent series can be generated by two lookup tables indicating what bit follows which $\mu(t)$. Whatever bit is indicated by the first table, the opposite bit is indicated by the second. Whenever an entry in a table is used for a given $\mu(t)$, the other table is used when $\mu(t)$ occurs again [14]. These tables are identical to strategy pairs at the maximum Hamming distance in the MG. No matter which of the $2^{m+1}$=16 possible strategies is used for the first table, and regardless of which of the $2^m$=4 possible $\mu(t)$ are used to initiate it, the time series generated by these tables will rapidly settle into perfect anti-persistence.

The "persistence" P of a given series at scale $m_s$ is thus simply the proportion of persistent such following bits, counting every instance of each of the $2^m$ possible histories. Its "anti-persistence" is 1 minus the persistence. Other waya of stating the same thing is that given $m_s$, the persistence of a series is the proportion of times that histories of length $m_s + 1$ end in bits 00 or 11; anti-persistence is the proportion they end in 01 or 10

The process generating a given empirical series may ne unknown. This unknown process may itself be a memory-related process such as in the games we are discussing; it need not be (it could be, for example, completely random). The process may likewise be Markovian and memory-related as are the time-series generated by the TH games; it may be memory-related but non-Markovian as the non-TH version of these games. If the process is memory-related, whether

Markovian or not, we need to distinguish between the unknown length $m$ (or $m+\tau$) underlying the process and a length we denote as $m_s$ indicating the scale of our analysis. Intuitively, it would seem that choosing $m_s = m$ or $m_s = m+\tau$, would yield the most robust analysis of persistence versus anti-persistence. But if the memory length of the process is unknown, this cannot be done. In the case of a TH game, all paths of length $m+\tau$ transition to other paths of equal length with known probabilities as these games are Markovian. The scale $m+\tau$ would seem even more natural since all quantities can be determined exactly using analytic methods, at least in principle. See [15,16] for an illuminating study on how to determine the optimal coarse-grained scale in simple cellular automata.

However, for m or $\tau$ large, the transition matrices become intractably large as well, scaling as $(m+\tau)^2$. We thus need to know whether the degree of persistence/antipersistence may be approximated at a lower effective $m_s$: I.e., given a binary time-series generated by an unknown process, may we usefully characterize its persistence by a small value of $m_s$ to replace its 'actual' $m$ or $m+\tau$? Before analyzing the degree of persistence and anti-persistence in the MG, MAJG and $G, we first show that, in fact, relatively small values of $m_s$ do successfully characterize persistence.

We implement an algorithm to characterize the persistence of a binary time series as described above. We find sharp differences in the degree of persistence between the time series generated by the MG on the one hand and the time series generated by the MAJG and $G on the other. A less sharp distinction also emerges between the MAJG and the $G. We find as well that characteristic distinctions arise at all reasonable m, attenuating as m grows large and at all reasonable scales. This last point is important: While the degree of persistence is obviously best captured for the TH variants of these games as the natural scale $m+\tau$ (since the TH games are Markovian), it is not obvious that a small-m scale will effectively capture distinctions in the MG, MAJG and $G proper as the natural scale is large and unbounded. It emerges that in general, if a significant degree of persistence or antipersistence is characteristic at a large-m scale, it may be approximated by a low-m analysis. We demonstrate this, and the differential characteristics of the respective time series in the following.

**Figure 8** illustrates graphically the mean degree of persistence or anti-persistence averaged over 25 different initializations identically shared by each of the THMG, THMAJG and TH$G with N=31, S=2, $\tau$=100 and $m \in \{2,3,...,10\}$, scale $\in \{2,3,...10\}$. The generally darker shade in all squares of the grid representing the THMG implies values closer to or consistent with anti-persistence (<0.5, with 0.5 representing the equal degree of persistence and anti-persistence of a random sequence), those in the THMAJG and TH$G with persistence (>0.5).

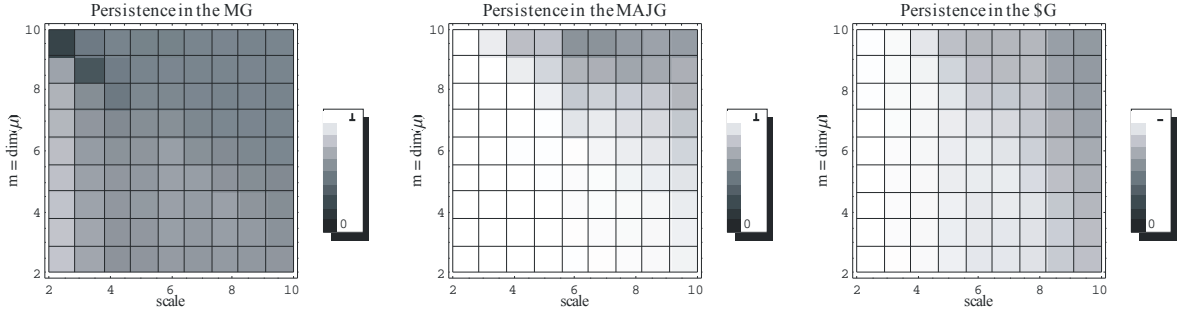

**Figure 8:** Persistence (white)/Anti-Persistence (black) at various scales and memory lengths in the MG, MAJG and $G. The grey scale between 0 and 1 given to the right of the checkboards encodes the degree of persistence at the chosen scale (abscissa) for different m values (ordinate), calculated as described in the text, using game-generated binary histories of length 1000 over 100 different runs for each game type

The fraction of persistent sequences up cells of increasing m are roughly similar on a relative basis up the columns of different scales, but shifted toward the random, especially for the THMAJG and TH$G. The fraction of persistent sequences up cells of increasing m in the THMG shows a shifting transition point. This feature is seen most sharply along the upward-and-to-the-right diagonal which represents the relation scale=m. Figure 9 charts the degree of persistence along this diagonal for all three games.

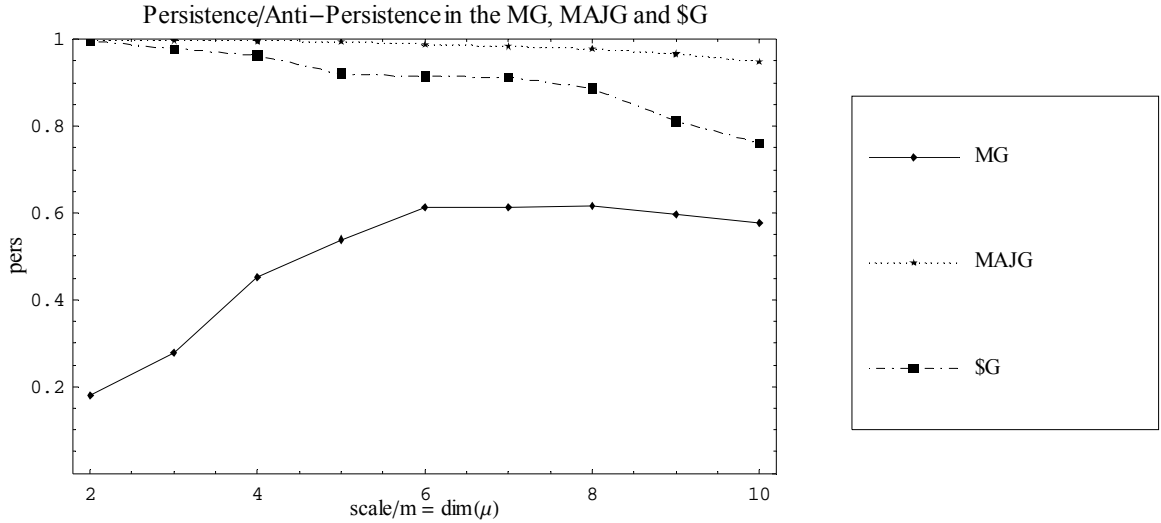

**Figure 9**: Persistence/Anti-Persistence at various scales = memory lengths in the MG, MAJG and $G

At the phase transition ($m_c = 4$ for N=31), the time-series generated by the THMG when the scale equals m undergoes a transition from anti-persistence to persistence and then declines asymptotically to the random limit 0.5. When $m_s \neq m$ this transition occurs at either smaller or larger values of m. Both the THMAJG and TH$G generate persistent time-series exclusively. The degree of persistence declines monotonically to the random limit 0.5 with increasing m. Persistence in the THMAJG is always greater than in the TH$G.

## 6 Conclusions

The "illusion of control" is an unfortunate confounding effect that appears in many situations of "bounded rationality" where optimization occurs with limited (inadequate) information. However ubiquitous the illusion may be, it is not universal. In ref. [2], we note that the illusion of control effect in the THMG is fundamentally due to three ingredients: (i) the minority mechanism (an agent or a strategy gains when in the minority and loses otherwise); (ii) the selection of strategies by many agents because they were previously in the minority, hence less likely to be so in the present; and (iii) the crowding of strategies (i.e., few strategies for many agents). We see in the preceding analysis of persistence, that there is a close relationship among these three characteristics, a high degree of anti-persistence in the resulting time-series and the illusion of control. On the other hand, genuine control is more likely to be present when the underlying mechanism employed by agents is not of the minority type (as in the MAJG and $G) and the resulting time-series is therefore more likely to be persistent. In another paper [3], we extend this analysis to the types of Hamiltonian cycles on graphs found associated with persistent and anti-persistent series, and employ these methods for generating predictors of empirically generated time-series, both in models and in the real world.

## Acknowledgements

We are grateful to Damien Challet who initially posed the question that resulted in this manuscript, as to whether an "illusion of control" discovered in [2] for Minority Games would be found in Majority Games as well.

## Appendix: Analytic Methods for the THMG, THMAJG and TH$G for choosing the best strategy

To emphasize the relation of the THMG, THMAJG and TH$G to market-games and to either genuine or illusory optimization, we have transformed the fundamental results of the games from statements on the properties of $\sigma^2/N$ to change in wealth, i.e., $\langle \Delta W/\Delta t \rangle$ for agents and $\langle \Delta W/\Delta t \rangle$ for strategies. We use the simplest possible formulation—if an agent's actual (or strategy's hypothetical) vote places it in the winning group (minority in the THMG, majority in the THMAJG and TH$G), it scores +1 points, otherwise −1. Formally: At every discrete time-step $t$, each agent independently re-selects one of its $S$ strategies. It "votes" as the selected strategy dictates by taking one of two "actions," designated by a binary value:

$$a_i(t) \in \{1,0\}, \ \forall \ i,t \quad (A.1)$$

In the simplest version of the MG, MAJG and $G with $N$ agents, every agent has $S = 2$ strategies and $m = 2$. In the THMG, THMAJG and TH$G, the point (or score) table associated with strategies is not maintained from the beginning of the game and is not ever growing. It is a rolling window of finite length $\tau$ (in the simplest case $\tau = 1$). The standard MG reaches an equilibrium state after a finite number of steps $t_{st}$. At this point, the dynamics and the behavior of individual agents for a given initial quenched disorder in the MG are indistinguishable from an otherwise identical THMG with $\tau \geq t_{st}$. The present formulation follows closely that presented in Appendix A.1 of [2], applying it simultaneously to all three games, the THMG, THMAJG and TH$G.

The state of the system as a whole at time $t$ is a mapping of the sum of all the agents' actions to the integer set $\{2N_1 - N\}$, where $N_1$ is the number of 1 votes and $N_0 = N - N_1$. This mapping is defined as:

$$A(t) = 2\sum_{i=1}^{N} a_i(t) - N = N_1 - N_0 \qquad (A.2)$$

If $A(t) > \frac{N}{2}$, then the minority of agents will have chosen 0 at time $t$ ($N_0 < N_1$); if $A(t) < \frac{N}{2}$, then the minority of agents will have chosen 1 at time $t$ ($N_1 < N_0$). In the MG the minority choice is the "winning" decision for $t$. In both the MAJG and \$G the majority choice is the "winning" decision for $t$. This choice is mapped back to $\{0,1\}$:

$$D_{sys}(t) = -\text{Sgn}[A(t)] \quad \therefore D_{sys}(t) \in \{-1,+1\} \to \{0,1\} \qquad (A.3)$$

For the non-TH version all three games, binary strings of length $m$ form histories $\mu(t)$, with $m = \dim[\mu(t)]$. For the TH games, binary strings of length $m+\tau$ form paths (or "path histories"), with $m+\tau = \dim(\mu_t)$, where we define $\mu(t)$ as a history in the standard MG and $\mu_t$ as a path in the THMG. Note that for memory m, there are $2^{2^m}$ possible strategies from which agents select S at random. However as first detailed in ref. [11], the space of strategies can be minimally spanned by a subset of all possible strategies. This reduced strategy space [RSS] has dimension $2^{m+1}$. As in Ref. [10], we may represent this quenched disorder in the allocation of strategies among agents (from the RSS) by a $\dim = \prod_{s=1}^{S} 2^{m+1}$ tensor, $\hat{\Omega}$ (or from the full strategy space by a $\dim = \prod_{s=1}^{S} 2^{2^m}$ tensor). The $2^{m+1}$ (or $2^{2^m}$) strategies are arranged in numerical order along the edges of $\hat{\Omega}$. Each entry represents the number of agents with the set of strategies indicated by the element's position. Then as demonstrated in [10], any THMG has a Markov chain formulation. For $\{m, S, N\} = \{2, 2, 31\}$ and using the RSS, the typical initial quenched disorder in the strategies attributed to each of the $N$ agents is represented by an $8 \times 8$ matrix $\hat{\Omega}$ and its symmetrized equivalent $\hat{\Psi} = \frac{1}{2}(\hat{\Omega} + \hat{\Omega}^T)$. Positions along all $S$ edges of $\hat{\Omega}$ represent an ordered listing of all available strategies. The numerical values $\Omega_{ij...}$ in $\hat{\Omega}$ indicate the number of times a specific strategy-tuple has been selected. (E.g., for two strategies per agent, $S=2$, $\Omega_{2,5}=3$ means that there are 3 agents with strategy 2 and strategy 5.) Without loss of generality, we may express $\hat{\Omega}$ in upper-triangular form since the order of strategies in a agent has no meaning. The example (A.4) is a typical such tensor $\hat{\Omega}$ for $S = 2, N = 31$.

$$\hat{\Omega} = \begin{pmatrix} 1 & 2 & 0 & 0 & 1 & 1 & 0 & 0 \\ 0 & 0 & 0 & 0 & 3 & 3 & 1 & 1 \\ 0 & 0 & 2 & 0 & 1 & 0 & 0 & 0 \\ 0 & 0 & 0 & 1 & 1 & 0 & 0 & 1 \\ 0 & 0 & 0 & 0 & 1 & 0 & 2 & 1 \\ 0 & 0 & 0 & 0 & 0 & 2 & 2 & 1 \\ 0 & 0 & 0 & 0 & 0 & 0 & 2 & 1 \\ 0 & 0 & 0 & 0 & 0 & 0 & 0 & 0 \end{pmatrix} \qquad (A.4)$$

Actions are drawn from a reduced strategy space (RSS) of dimension $r = 2^{m+1}$. Each action is associated with a strategy $k$ and a history $\mu(t)$. There are $2^m$ histories. Together, the score for all actions by history and strategy can be represented in table form as a $\dim(\text{RSS}) \times 2^m$ binary matrix i.e., score($a_k^{\mu_t}$) $\in \{-1, +1\}$. In this case, the table reads:

$$\hat{a} \equiv \begin{pmatrix} -1 & -1 & -1 & -1 \\ -1 & -1 & +1 & +1 \\ -1 & +1 & -1 & +1 \\ -1 & +1 & +1 & -1 \\ +1 & -1 & -1 & +1 \\ +1 & -1 & +1 & -1 \\ +1 & +1 & -1 & -1 \\ +1 & +1 & +1 & +1 \end{pmatrix} \qquad (A.5)$$

The change in wealth (point gain or loss) associated with each of the r = 8 strategies for the 8 allowed transitions among the 4 histories) at any time $t$ for each of the three games is then:

$$\delta \vec{S}^{\min}_{\mu(t), \mu(t-1)} = +\left(\hat{a}^T\right)_{\mu(t)} \times \{2 Mod[\mu(t-1), 2] - 1\} \qquad (A.6)$$

$$\delta \vec{S}^{maj}_{\mu(t), \mu(t-1)} = -\left(\hat{a}^T\right)_{\mu(t)} \times \{2 Mod[\mu(t-1), 2] - 1\} \qquad (A.7)$$

$$\delta \vec{S}^{\$}_{\mu(t), \mu(t-1)} = -\left(\hat{a}^T\right)_{\mu(t-1)} \times \{2 Mod[\mu(t-1), 2] - 1\} \qquad (A.8)$$

$Mod[x, y]$ is "x modulo y"; $\mu(t)$ and $\mu(t-1)$ label each of the 4 histories $\{00, 01, 10, 11\}$ hence take on one of values $\{1, 2, 3, 4\}$. Equations (A.6), (A.7) and (A.8) pick out from (A.5) the correct change in wealth over a single step since the strategies are ordered in symmetrical sequence.

The change in points associated with each strategy for each of the allowed transitions along all the $\tau$ histories (i.e., along the path $\mu_t$, accounting for the last $\tau$ time steps used to score the strategies) is:

$$\vec{s}_{\mu_t}^{\,game} \equiv \sum_{i=0}^{\tau-1} \delta \vec{S}_{\mu(t-i),\mu(t-i-1)}^{\,game} \tag{A.9}$$

(A.9) accounts for the change in points along path $\mu_t$ by summing them over all transitions on the path, with $game \in \{min, maj, \$\}$ For example, for $m = 2$ and $\tau = 1$, the strategy scores are kept for only a single time-step and $\tau - 1 = 0$ so the sum vanishes. (A.9) in matrix form therefore reduces to the score:

$$\vec{s}_{\mu_t}^{\,game} \equiv \delta \vec{S}_{\mu(t),\mu(t-1)}^{\,game} \tag{A.10}$$

or, listing the results for all 8 path histories:

$$\hat{\mathbf{s}}_{\mu}^{\,game} \equiv \delta \hat{\mathbf{S}}^{\,game} \tag{A.11}$$

$\delta \hat{\mathbf{S}}^{\,game}$ is an 8×8 matrix that can be read as a lookup table. It denotes the change in points accumulated over $\tau = 1$ time steps for each of the 8 strategies over each of the 8 path-histories.

Instead of computing $A^{game}(t)$, we compute $A^{game}(\mu_t)$. Then for each of the $2^{m+\tau} = 8$ possible $\mu_t$, $A(\mu_t)$ is composed of a subset of wholly determined agent votes and a subset of undetermined agents whose votes must be determined by a coin toss:

$$A^{game}(\mu_t) = A_D^{game}(\mu_t) + A_U^{game}(\mu_t) \tag{A.12}$$

Some agents are undetermined at time t because their strategies have the same score and the tie has to be broken with a coin toss. $A_U^{game}(\mu_t)$ is a random variable characterized by the binomial distribution. Its actual value varies with the number of undetermined agents $N_U^{game}$. This number can be explicated (using an extension to the method employed in [10] for the THMG) as:

$$N_U^{min}(\mu_t) =$$
$$\left\{ \left(1 - \left[(\hat{a}^\mathsf{T})_{(\text{Mod}[\mu_t-1,4]+1)} \otimes_\delta (\hat{a}^\mathsf{T})_{(\text{Mod}[\mu_t-1,4]+1)}\right]\right) \circ \left(\vec{s}_{\mu_t}^{\,min} \otimes_\delta \vec{s}_{\mu_t}^{\,min}\right) \circ \hat{\mathbf{\Omega}} \right\}_{(\text{Mod}[\mu_t-1,2^m]+1)} \tag{A.13}$$

$$N_U^{maj}(\mu_t) =$$
$$\left\{ \left(1 - \left[(\hat{a}^\mathsf{T})_{(\text{Mod}[\mu_t-1,4]+1)} \otimes_\delta (\hat{a}^\mathsf{T})_{(\text{Mod}[\mu_t-1,4]+1)}\right]\right) \circ \left(\vec{s}_{\mu_t}^{\,maj} \otimes_\delta \vec{s}_{\mu_t}^{\,maj}\right) \circ \hat{\mathbf{\Omega}} \right\}_{(\text{Mod}[\mu_t-1,2^m]+1)} \tag{A.14}$$

$$N_U^{\$}(\mu_t) =$$
$$\left\{ \left(1 - \left[(\hat{a}^\mathsf{T})_{(\text{Mod}[\mu_{t-1}-1,4]+1)} \otimes_\delta (\hat{a}^\mathsf{T})_{(\text{Mod}[\mu_{t-1}-1,4]+1)}\right]\right) \circ \left(\vec{s}_{\mu_t}^{\,\$} \otimes_\delta \vec{s}_{\mu_t}^{\,\$}\right) \circ \hat{\mathbf{\Omega}} \right\}_{(\text{Mod}[\mu_{t-1}-1,2^m]+1)} \tag{A.15}$$

"$\otimes_\delta$" is a generalized outer product, with the product being the Kronecker delta. $\vec{N}_U$ constitutes a vector of such values. The summed value of all undetermined decisions for a given $\mu_t$ is distributed binomially. Note that (A.13) and (A.14) are structurally identical while (A.15) differs from these in that the indices on $(\hat{a}^T)$ and on the entire expression reference path-histories $\mu_{t-1}$ rather than $\mu_t$, reflecting the one-step time-lag in the payoff for the \$G. Similarly:

$$A_D^{min}(\mu_t) = \left(\sum_{r=1}^{8}\left\{\left[\left(1-Sgn\left[\vec{s}_{\mu_t}^{min} \ominus \vec{s}_{\mu_t}^{min}\right]\right) \circ \hat{\Psi}\right] \bullet \hat{a}\right\}_r\right)_{(Mod[\mu_t-1, 2^m]+1)} \quad (A.16)$$

$$A_D^{maj}(\mu_t) = \left(\sum_{r=1}^{8}\left\{\left[\left(1-Sgn\left[\vec{s}_{\mu_t}^{maj} \ominus \vec{s}_{\mu_t}^{maj}\right]\right) \circ \hat{\Psi}\right] \bullet (-\hat{a})\right\}_r\right)_{(Mod[\mu_t-1, 2^m]+1)} \quad (A.17)$$

$$A_D^{\$}(\mu_t) = \left(\sum_{r=1}^{8}\left\{\left[\left(1-Sgn\left[\vec{s}_{\mu_t}^{\$} \ominus \vec{s}_{\mu_t}^{\$}\right]\right) \circ \hat{\Psi}\right] \bullet (-\hat{a})\right\}_r\right)_{(Mod[\mu_{t-1}-1, 2^m]+1)} \quad (A.18)$$

Details of the derivation as applying to the THMG may also be found in Ref. [17]. We define $\bar{A}_D$ as a vector of the determined contributions to $A(t)$ for each path $\mu_t$. In expressions (A.11) and (A.12) $\mu_t$ numbers paths from 1 to 8 and is therefore here an index. $\vec{s}_{\mu_t}$ is the "$\mu_t^{th}$" vector of net point gains or losses for each strategy when at $t$ the system has traversed the path $\mu_t$ ( i.e., it is the "$\mu_t^{th}$" element of the matrix $\hat{s}_\mu = \delta\hat{S}$ in (A.11)). "$\ominus$" is a generalized outer product of two vectors with subtraction as the product. The two vectors in this instance are the same, i.e., $\vec{s}_{\mu_t}$. "$\circ$" is Hadamard (element-by-element) multiplication and "$\bullet$" the standard inner product. The index $r$ refers to strategies in the RSS. Summation over $r$ transforms the base-ten code for $\mu_t$ into $\{1,2,3,4,1,2,3,4\}$. Selection of the proper number is indicated by the subscript expression on the entire right-hand side of (A.13). This expression yields an index number, i.e., selection takes place 1 + Modulo 4 with respect to the value of $(\mu_t-1)$ for the THMG and with respect to the value of $(\mu_{t-1}-1)$ for the THMAJG and TH\$G.

To obtain the transition matrix for the system as a whole, we require the $2^{m+\tau} \times 2^{m+\tau}$ adjacency matrix that filters out disallowed transitions. Its elements are $\Gamma_{\mu_t,\mu_{t-1}}$:

$$\hat{\mathbf{\Gamma}} = \begin{pmatrix} 1 & 0 & 0 & 0 & 1 & 0 & 0 & 0 \\ 1 & 0 & 0 & 0 & 1 & 0 & 0 & 0 \\ 0 & 1 & 0 & 0 & 0 & 1 & 0 & 0 \\ 0 & 1 & 0 & 0 & 0 & 1 & 0 & 0 \\ 0 & 0 & 1 & 0 & 0 & 0 & 1 & 0 \\ 0 & 0 & 1 & 0 & 0 & 0 & 1 & 0 \\ 0 & 0 & 0 & 1 & 0 & 0 & 0 & 1 \\ 0 & 0 & 0 & 1 & 0 & 0 & 0 & 1 \end{pmatrix} \quad (A.19)$$

Equations (A.13), (A.14) and (A.15) yield the history-dependent $(m+\tau) \times (m+\tau)$ matrix $\hat{\mathbf{T}}$ with elements $T_{\mu_t,\mu_{t-1}}$, representing the 16 allowed probabilities of transitions between the two sets of 8 path-histories $\mu_t$ and $\mu_{t-1}$ (the game-type superscripts on $A_D$ and $N_U$ are understood in context):

$$T_{\mu_t,\mu_{t-1}}^{game} = \Gamma_{\mu_t,\mu_{t-1}}^{game} \times$$

$$\sum_{x=0}^{N_U^{game}(\mu_t)} \left\{ \mathbb{C}_x^{N_U^{game}(\mu_t)} \left(\frac{1}{2}\right)^{N_U^{game}(\mu_t)} \times \delta\left[\text{Sgn}\left(A_D^{game}(\mu_t) + 2x - N_U^{game}(\mu_t)\right) + (2\text{Mod}\{\mu_{t-1},2\} - 1)\right] \right\} \quad (A.20)$$

The C expression $\mathbb{C}_x^{N_U^{game}(\mu_t)} \left(\frac{1}{2}\right)^{N_U^{game}(\mu_t)}$ in (A.20) represents the binomial distribution of undetermined outcomes under a fair coin-toss with mean = $A_D^{game}(\mu_t)$. Given a specific $\hat{\mathbf{\Omega}}$,

$$\langle A^{game}(\mu_t) \rangle = A_D^{game}(\mu_t) \; \forall \; \mu_t \quad (A.21)$$

We now tabulate the number of times each strategy is represented in $\hat{\mathbf{\Omega}}$, regardless of coupling (i.e., of which strategies are associated in forming agent S-tuples), a value which is the same across all games (hence we drop the "game" superscript):

$$\bar{\kappa} \equiv \sum_{k=1}^{2^{m+\tau}} \left(\hat{\mathbf{\Omega}} + \mathbf{\Omega}^\top\right)_k = 2\sum_{k=1}^{2^{m+\tau}} \hat{\mathbf{\Psi}}_k = \left\{n(\sigma_1), n(\sigma_2), \ldots n(\sigma_{2^{m+\tau}})\right\} \quad (A.22)$$

where $\sigma_k$ is the $k^{\text{th}}$ strategy in the RSS, $\hat{\mathbf{\Omega}}_k, \hat{\mathbf{\Omega}}_k^\top$ and $\hat{\mathbf{\Psi}}_k$ are the $k^{\text{th}}$ element (vector) in each tensor and $n(\sigma_k)$ represents the number of times $\sigma_k$ is present across all strategy tuples. Therefore

$$\langle \Delta W_{Agent} \rangle = \pm \frac{1}{N} Abs(\bar{A}_D) \cdot \bar{\mu} \quad (A.23)$$

(with the minus sign for the MG, otherwise not, i.e., the awarding of points is the negative of the direction of the vote imbalance for the MG, and in the direction of the imbalance in the MAJG and \$G.) and

$$\left\langle \Delta W_{Strategy} \right\rangle = \tfrac{1}{2N}\left(\hat{\mathbf{s}}_\mu \cdot \vec{\kappa}\right) \cdot \vec{\mu} \tag{A.24}$$

with $\vec{\mu}$ the normalized steady-state probability vector for $\hat{\mathbf{T}}$. Expression (A.23) states that the mean per-step change in wealth for agents equals ±1 times the probability-weighted sum of the (absolute value of the) *determined* vote imbalance associated with a given history, with a minus sign for the THMG. Expression (A.24) states that the mean per-step change in wealth for individual strategies equals the probability-weighted sum of the representation of each strategy (in a given $\hat{\mathbf{\Omega}}$) times the sum over the per-step wealth change associated with every history.